\documentclass[10pt,english]{article}
\usepackage{mathptmx}
\usepackage{helvet}
\usepackage{courier}
\usepackage[T1]{fontenc}
\usepackage[latin9]{inputenc}
\usepackage[a4paper]{geometry}
\usepackage{float}
\usepackage{amsmath}
\usepackage{graphicx}
\usepackage{setspace}
\usepackage{nameref}

\makeatletter

\def\d{\mathrm{d}}

\def\Grad{\vec{\mathrm{gra}}\mathrm{d}}

\def\Div{\mathrm{div}}

\def\SH{{\textstyle\frac{1}{2}}}

\def\degree{\ifmmode ^o\else $^o$\fi}

\def\gammabar{\hbox to 0cm{\hglue 0.2ex--}\gamma}

\usepackage[scanall]{psfrag}

\def\legende#1{\raise -0.5ex\hbox{\smash{#1}}}

\def\pflegende#1[#2]#3{\@pfl{#1}#2{#3}}
\def\@pfl#1#2#3#4{%
  \if#3l%
    \if#2t%
      \psfrag{#1}[Bl][Bl]{\raise -0.5ex\hbox{\vtop{\@pfl@l{#4}}}}%
    \else\if#2c%
      \psfrag{#1}[cl][Bl]{\hbox{\vtop{\@pfl@l{#4}}}}%
    \else
      \psfrag{#1}[Bl][Bl]{\raise -0.5ex\hbox{\vbox{\@pfl@l{#4}}}}%
    \fi\fi
  \else\if#3c%
    \if#2t%
      \psfrag{#1}[Bc][Bc]{\raise -0.5ex\hbox{\vtop{\@pfl@c{#4}}}}%
    \else\if#2c%
      \psfrag{#1}[cc][Bc]{\hbox{\vtop{\@pfl@c{#4}}}}%
    \else
      \psfrag{#1}[Bc][Bc]{\raise -0.5ex\hbox{\vbox{\@pfl@c{#4}}}}%
    \fi\fi
  \else
    \if#2t%
      \psfrag{#1}[Br][Br]{\raise -0.5ex\hbox{\vtop{\@pfl@r{#4}}}}%
    \else\if#2c%
      \psfrag{#1}[cr][Br]{\hbox{\vtop{\@pfl@r{#4}}}}%
    \else
      \psfrag{#1}[Br][Br]{\raise -0.5ex\hbox{\vbox{\@pfl@r{#4}}}}%
    \fi\fi
  \fi\fi
}
\def\@pfl@l#1{\leftskip=0pt plus 1cm\parfillskip=0pt\baselineskip2ex#1}
\def\@pfl@c#1{\leftskip=0pt plus 1cm\rightskip=\leftskip\parfillskip=0pt\baselineskip2ex#1}
\def\@pfl@r#1{\rightskip=0pt plus 1cm\parfillskip=0pt\baselineskip2ex#1}

\newif\ifafficherappelsfig
\afficherappelsfigtrue

\def\deplace#1#2#3{\hglue 0pt\vbox to 0pt{\vglue #2\hbox to 0pt{\hglue #1\smash{#3}}}}

\def\figfont{\small}
\def\Fig#1#2{%
  Figure #1, J. Lucas, C. Margo, Y. Oussar and S. Holé\par%
  #2\vglue 4cm\par%
}

\makeatother

\usepackage{babel}
\begin{document}

\title{Spatial Resolution in Electrical Capacitance Tomography}

\author{Jérôme Lucas, Cédric Margo, Yacine Oussar and Stéphane Holé%
\thanks{Corresponding author: stephane.hole@espci.fr%
}\medskip\\
{\small{}Laboratoire de Physique et d'Étude des Matériaux (LPEM)}\\
{\small{}PSL Research University, ESPCI-ParisTech -- Sorbonne Universités,
UPMC Univ Paris 06 -- CNRS, UMR8213}\\
{\small{}10, rue Vauquelin -- 75005 Paris -- France}}
\maketitle
\begin{abstract}
Electrical Capacitance Tomography (ECT) is an imaging technique providing
the distribution of permittivity in a medium by the mean of electrodes.
As for any imaging systems, the reachable spatial resolution is a
key parameter. In this paper the spatial resolution of ECT sensors
is analyzed in terms of the accuracy of an object position and of
the ability to distinguish between two close objects. Cylindrical
geometry sensors are particularly studied and the example of a square
geometry sensor is used to show how to study any other geometries.
In cylindrical geometries, it is shown that a 50\% gap between electrodes
is a good compromise and that increasing the number of electrodes
improves the spatial resolution near the electrodes but decreases
the spatial resolution in the center. The best spatial resolution
at the center of the sensor is obtained with 3 or 4 electrodes. In
the square geometry studied, it is shown that a better distribution
of the spatial resolution is obtained when there are electrodes in
the corners.
\end{abstract}
\textbf{keywords:} Electrical Capacitance Tomography, capacitive sensor,
spatial resolution

\section{Introduction}

Capacitive sensors are very convenient because they only consist of
electrodes and are sensitive to the electrical properties of materials
and to their distribution. Moreover they can work at low frequency
with low power consumption. Capacitive sensors are used in many applications,
for instance proximity detection, material characterization, hygrometry
measurement, position control and fraction measurement \cite{larry1997,kolbRSI1998,RittersmaSAA2002,HarreySAB2002,BadeltBRM2008,HuSR2010,ElbukenSAA2011,MargoISMTII2011,Sheldon2012,HudspethIAAPMS2012}.
In that latter case, it is necessary to estimate the volume fraction
of a given material in a host material, for instance water or air
in oil \cite{GeraetsIJMF1988,WangMST2009,MargoESA2012}. It is quite
a complex sensor since it is necessary to ensure that the fraction
of added material will give the same signal whatever its position
in the sensor sensitive volume \cite{CuvignySFE2010,MargoISMTII2011}.
That can be achieved by choosing a uniform sensitivity map sensor,
but it is not sufficient if the permittivity of the added material
varies since a small fraction of large permittivity material, a water
drop for instance, would yield the same signal as a larger fraction
of a lower permittivity material, an air bubble for instance.

Electrical Capacitance Tomography (ECT) \cite{YangMST2010,NeumayerI2MTC2012}
is a technique for estimating the permittivity distribution in the
sensor volume. It usually consists of electrodes evenly distributed
around a cylinder \cite{YangMST2010}. The capacitance between the
electrodes are measured and, because capacitance directly depends
on permittivity, the application of an appropriate algorithm \cite{BadeltBRM2008}
to the measurements makes it possible to estimate the permittivity
distribution.

ECT sensors are indeed capacitive sensors and therefore obey to electrostatic
laws \cite{larry1997}. As the effect of electric field depends on
the distance to the electrodes, spatial resolution should depend on
the position in the sensor sensitive volume. As a consequence the
number of electrodes, their shape and the mesh used in the reconstruction
algorithm could be optimized as to converge quicker or lead to better
results.

In this paper, the physical background of ECT is described in the
first section. Sensitivity map is analytically calculated and Landweber
algorithm is recalled as an illustration. Spatial resolution is discussed
in the second section. It is applied to ECT systems particularly for
the localization of a material and for the discrimination between
two materials. The influence of (i) the size and position of materials
to detect, of (ii) the sensor diameter, and of (iii) the number and
size of the electrodes, on spatial resolution is studied for a cylindrical
geometry. A square geometry is also used to illustrate the design
choices that can be taken in more complex geometries.

\section{Physical background of ECT}

An Electrical Capacitance Tomography (ECT) system is a multi-electrode
capacitive sensor for estimating the permittivity distribution of
materials located in the sensor sensitive volume. It is directly based
on the influence of a material permittivity on the capacitance between
the electrodes of the sensor. The charges on the electrodes of any
capacitive sensor depend on the voltage applied to each electrode
and is given in tensorial notation by
\begin{equation}
Q_{i}=c_{ij}V_{j},\label{eq:CapacitiveTensor1}
\end{equation}
where $Q_{i}$ is the charge quantity on electrode $i$, $c_{ij}$
is the capacitive tensor and $V_{j}$ is the voltage applied to electrode
$j$. Notice that in tensorial notation, a sum is insinuated over
all values of any subscript that appears twice in a term. In (\ref{eq:CapacitiveTensor1}),
there is an insinuated sum over all values of $j$, that is to say
over all electrodes. 

When $i\neq j$, the capacitive tensor coefficient $c_{ij}$ depends
on the capacitance $C_{ij}$ between electrodes $i$ and $j$ (see
Figure~\ref{fig:GeneralCapacitiveSensor}). Since $c_{ij}$ links
the charges on electrode $i$ to the voltage applied to electrode
$j$ when all other electrodes are grounded, one has $c_{ij}=-C_{ij}$.
One can verify that capacitive tensor is symmetrical, that is to say
$c_{ij}=c_{ji}$. When $i=j$ the capacitive tensor coefficient $c_{ii}$
links the charges on electrode $i$ to the voltage on the same electrode
when all other electrodes are grounded. As a consequence coefficient
$c_{ii}=\sum_{j}C_{ij}$ is the sum of the capacitance of all capacitors
connected to electrode $i$. Finally the capacitive tensor $c_{ij}$
can be expressed from the capacitors connecting electrodes together
as
\begin{equation}
c_{ij}=\begin{bmatrix}\sum_{j}C_{1j} & -C_{12} & -C_{13} & \cdots & -C_{1N}\\
-C_{12} & \sum_{j}C_{2j} & -C_{23} & \cdots & -C_{2N}\\
-C_{13} & -C_{23} & \sum_{j}C_{3j} & \cdots & -C_{3N}\\
\vdots & \vdots & \vdots & \ddots & \vdots\\
-C_{1N} & -C_{2N} & -C_{3N} & \dots & \sum_{j}C_{Nj}
\end{bmatrix}.\label{eq:CapacitiveTensor2}
\end{equation}
For a $N$-electrode sensor there are at most $N(N-1)/2$ different
capacitors and thus $N(N-1)/2$ independent data can be obtained at
most, for instance permittivity at $N(N-1)/2$ positions.

\subsection{Capacitive sensors}

A variation in the environment of a capacitive sensor results in a
signal. For a measurement electrode $i$ the variation of (\ref{eq:CapacitiveTensor1})
leads to
\begin{equation}
\delta Q_{i}=\delta\left(c_{ij}V_{j}\right)=\delta c_{ij}V_{j}+c_{ij}\delta V_{j\neq i}+c_{ii}\delta V_{i}.\label{eq:SignalVariation}
\end{equation}

In short-circuit measurement conditions, $\delta V_{i}=0$ and the
signal is the variation of charges $\delta Q_{i}$ on electrode $i$.
In open-circuit measurement conditions, $\delta Q_{i}=0$ and the
signal is the variation of voltage $\delta V_{i}$ on electrode $i$.
In both cases the signal results from a voltage variation $\delta V_{j}$
on the other electrodes ($i\neq j$) and/or from a capacitive tensor
variation $\delta c_{ij}$ due, for instance, to a local variation
of permittivity. Therefore both measurement conditions are connected
together. One has
\begin{equation}
\delta Q_{i}\equiv-c_{ii}\delta V_{i}.\label{eq:TheveninNorton}
\end{equation}
Capacitive tensor coefficients can be advantageously estimated by
energy considerations in the case of ECT sensors. The energy $W$
of an electrostatic system at equilibrium is the sum of the energy
held by each electrode, that is to say half the sum of the charge
quantity on each electrode multiplied by the voltage applied to that
electrode \cite{Griffiths1999}:
\begin{equation}
W=\SH Q_{i}V_{i}=\SH c_{ij}V_{i}V_{j}.\label{eq:Energy1}
\end{equation}
Energy $W$ can also be expressed as the integral over space of the
energy density $\SH\epsilon E^{2}$, where $\epsilon$ is the permittivity
and $\vec{E}$ is the electric field. One has 
\begin{equation}
W=\SH\int\epsilon\, E^{2}\,\d v.\label{eq:Energy2}
\end{equation}
Considering a $N$-electrode ECT sensor, the electric field $\vec{E}$
can be conveniently decomposed into the sum of the contribution of
each electrode by introducing $\vec{\xi}_{i}$, the electric field
produced by electrode $i$ when polarized to 1~V while all other
electrodes are grounded. One obtains 
\begin{equation}
\vec{E}=V_{i}\vec{\xi}_{i}\label{eq:ElectricFieldDecomposition}
\end{equation}
and thus (\ref{eq:Energy2}) becomes
\begin{equation}
W=\SH\int\epsilon\,(V_{i}\vec{\xi}_{i})\cdot(V_{j}\vec{\xi}_{j})\,\d v.\label{eq:Energy3}
\end{equation}
The identification between (\ref{eq:Energy1}) and (\ref{eq:Energy3})
leads finally to
\begin{equation}
c_{ij}=\int\epsilon\,\vec{\xi}_{i}\cdot\vec{\xi}_{j}\,\d v.\label{eq:CapacitiveTensorCoefficients}
\end{equation}

As far as capacitive sensor variation $\delta c_{ij}$ is concerned,
it has been demonstrated \cite{HolePRB2000,LucasMST2006} that the
variation of charges $\delta Q_{i}$ in short-circuit measurement
conditions resulting from a local permittivity variation $\delta\epsilon$
is at first order
\begin{equation}
\delta Q_{i}=\int\delta\epsilon\,\vec{E}\cdot\vec{\xi}_{i}\,\d v.\label{eq:PermittivitySensitivity}
\end{equation}
Using (\ref{eq:ElectricFieldDecomposition}) for the electric field,
one easily obtains $\delta c_{ij}$ by identification: 
\begin{equation}
\delta Q_{i}=V_{j}\,\int\delta\epsilon\,\vec{\xi}_{i}\cdot\vec{\xi}_{j}\,\d v=V_{j}\delta c_{ij}.\label{eq:CapacitiveTensorVariation}
\end{equation}
Field $\vec{\xi}_{i}$ can be seen as the influence of electrode $i$
in the capacitive system. Therefore it can be called the sensitivity
field of electrode $i$.

\subsection{ECT sensitivity matrix}

The reconstruction of the permittivity distribution in the sensor
sensitive volume depends on measurements and therefore relies on the
influence of permittivity in the signal. That influence is the signal
sensitivity matrix (tensor) $S_{ij}$ and is generally defined as
\begin{equation}
S_{ij}=\frac{1}{V}\times\frac{\partial m_{ij}}{\partial\epsilon},\label{eq:SensitivityMatrix1}
\end{equation}
where $m_{ij}$ denotes the measurement at electrode $i$ when electrode
$j$ is under the voltage $V$ while other electrodes are grounded.
In short-circuit measurement conditions, one has from (\ref{eq:SignalVariation})
and (\ref{eq:SensitivityMatrix1}) $S_{ij}=\delta c_{ij}/\delta\epsilon$.
Since $\epsilon$ depends on positions, it is convenient to decompose
the environment of the sensor into small elements of volume $\delta v_{k}$.
The permittivity is assumed uniform in each element $k$ and can vary
from $\epsilon$ to $\epsilon+\delta\epsilon^{\mathrm{max}}$. In
that case the following linear expression is often used \cite{XieIEEPG1992}:
\begin{equation}
S_{ijk}=\frac{\max(\delta v_{k})}{\delta v_{k}}\times\frac{c_{ij}(\epsilon+\delta\epsilon_{k}^{\mathrm{max}})-c_{ij}(\epsilon)}{\delta\epsilon^{\mathrm{max}}}.\label{eq:SensitivityMatrix2}
\end{equation}

It corresponds to the capacitance variation between electrode $i$
and $j$ when the permittivity in element $k$ goes from $\epsilon$
to $\epsilon+\delta\epsilon^{\mathrm{max}}$ while the permittivity
in other elements is $\epsilon$, normalized by the permittivity variation
$\delta\epsilon^{\mathrm{max}}$ and the volume $\delta v_{k}$ of
element $k$. From Equation (\ref{eq:CapacitiveTensorCoefficients})
and (\ref{eq:CapacitiveTensorVariation}), $S_{ijk}$ can be expressed
as
\begin{equation}
S_{ijk}=\frac{\max(\delta v_{k})}{\delta v_{k}}\times\int_{\delta v_{k}}\vec{\xi}_{i}\cdot\vec{\xi}_{j}\,\d v=\max(\delta v_{k})\times\left\langle \vec{\xi}_{i}\cdot\vec{\xi}_{j}\right\rangle _{k},\label{eq:SensitivityMatrix3}
\end{equation}
where $\left\langle \vec{\xi}_{i}\cdot\vec{\xi}_{j}\right\rangle _{k}$
is the mean value of $\vec{\xi}_{i}\cdot\vec{\xi}_{j}$ over element
$k$. The dot product $\vec{\xi}_{i}\cdot\vec{\xi}_{j}$ is therefore
the sensor sensibility density. If $\vec{x}_{k}$ refers to as the
barycenter of element $k$, one has 
\begin{equation}
\left\langle \vec{\xi}_{i}\cdot\vec{\xi}_{j}\right\rangle _{k}\approx\vec{\xi}_{i}(\vec{x}_{k})\cdot\vec{\xi}_{j}(\vec{x}_{k}).\label{eq:SensitivityDensity}
\end{equation}

When electrodes are evenly distributed, it is worth noting that sensitivity
matrix $S_{ijk}$ can be estimated with only one simulation and the
application of appropriate rotations by using (\ref{eq:SensitivityMatrix3})
and (\ref{eq:SensitivityDensity}) taking advantage of the symmetry
of cylindrical ECT systems.

\subsection{Analytic expression}

An analytic expression of the sensitivity matrix $S_{ijk}$ can be
useful in some situations. When $N$ electrodes are joined and evenly
distributed around a cylinder of radius $R$, as illustrated in Figure~\ref{fig:ECTSensor}a,
a conform transformation can be used for the sensitivity field: 
\begin{equation}
\tilde{\xi}_{0}=\xi_{0}^{x}-\imath\xi_{0}^{y}=-\frac{2R\,\sin(\pi/N)}{\pi\,(R^{2}-2Rz\,\cos(\pi/N)+z^{2})}.\label{eq:AnalyticSensitivityFieldJoined0}
\end{equation}

As when electrodes are joined, the sensitivity field $\vec{\xi}$
varies mostly near the electrode held to 1~V when electrodes are
separated. Thus the sensor geometry can be approached with the one
presented in Figure~\ref{fig:ECTSensor}b in which the ground electrode
is continuous instead of being separated. In that approached geometry,
the conform transformation giving the sensitivity field is:
\begin{eqnarray}
\tilde{\xi}_{0}=\xi_{0}^{x}-\imath\xi_{0}^{y} & = & \frac{\imath R\,\sqrt{2\cos(\pi/N)+2\cos(\eta\pi/N)}}{\pi\,(z+R)\,\sqrt{R^{2}\,\exp(-\imath\pi/N)-2Rz\,\cos(\eta\pi/N)+z^{2}\,\exp(\imath\pi/N)}}\nonumber \\
 &  & -\frac{\imath R\,\sqrt{2\cos(\pi/N)+2\cos(\eta\pi/N)}}{\pi\,(z+R)\,\sqrt{R^{2}\,\exp(\imath\pi/N)-2Rz\,\cos(\eta\pi/N)+z^{2}\,\exp(-\imath\pi/N)}}.\label{eq:AnalyticSensitivityFieldSeparated0}
\end{eqnarray}
In expression (\ref{eq:AnalyticSensitivityFieldSeparated0}), $\eta$
corresponds to the gap ratio between electrodes, a small gap corresponding
to a small coefficient $\eta$. In these expressions $z=x+\imath y$
is the coordinates in the complex plane ($\imath=\sqrt{-1}$) and
$\tilde{\xi}_{0}$ is the complex sensitivity field which real part
$\xi_{0}^{x}$ and imaginary part $\xi_{0}^{y}$ correspond respectively
to the $x$-component and the opposite of the $y$-component of the
sensitivity field when electrode $j=0$ is held to 1~V (thick in
the figure) while all other electrodes are grounded (thin in the figure).

By symmetry it is possible to obtain the field $\tilde{\xi}_{j}$
corresponding to any electrode numbered $j$. First the coordinate
of point $z$ has to be rotated by $-2\pi\, j/N$. Second the electric
field calculated with (\ref{eq:AnalyticSensitivityFieldJoined0})
or (\ref{eq:AnalyticSensitivityFieldSeparated0}) has also to be rotated
by $-2\pi\, j/N$. The rotation is in the same direction because the
imaginary part of $\tilde{\xi}$ is the opposite of the $y$-component
of the electric field. Mathematically speaking one has
\begin{equation}
\tilde{\xi}_{j}(z)=\exp(-2\imath\pi\, j/N)\times\tilde{\xi}_{0}(z\times\exp(-2\imath\pi\, j/N)).\label{eq:AnalyticSensitivityfieldJ}
\end{equation}
An analytic expression of the sensitivity matrix $S_{ijk}$ can therefore
be calculated with
\[
S_{ijk}=\max(\delta v_{k})\times\Re\left(\tilde{\xi}_{i}(z_{k})\times\overline{\tilde{\xi}_{j}(z_{k})}\right),
\]
where $\overline{Z}$ is the conjugate of the complex number $Z$
and $\Re(Z)$ is its real part. Figures \ref{fig:ECTSensitivityMapsJoined}
and \ref{fig:ECTSensitivityMapsDisjointed} shows the 4 principal
sensitivity maps of the 10-cm-radius 8-electrode cylindrical sensor
with joined or separated electrodes respectively. All other sensitivity
maps can be obtained by appropriate rotation or symmetry.

\subsection{Landweber algorithm}

The reconstruction of the permittivity distribution in the sensor
sensitive volume can be made by various algorithms \cite{NeumayerI2MTC2012,IsaksenMST1996,YangMST2003}.
Though Landweber algorithm is one of the slowest, it generally gives
the best results. That iterative algorithm is based on the optimization
of a criterion $J$ defined as
\begin{equation}
J=(m_{ij}-V\, S_{ijk}\delta\hat{\epsilon}_{k})^{2}.\label{eq:Criterion}
\end{equation}
Criterion $J$ is minimized when the estimated permittivity distribution
$\delta\hat{\epsilon}_{k}$ produces an estimated signal $V\, S_{ijk}\delta\hat{\epsilon}_{k}$
as close as possible to the real measurements $m_{ij}$. The iterative
algorithm consists then in the iterative minimization of criterion
$J$ by
\begin{equation}
\delta\hat{\epsilon}_{\ell}^{n+1}=\delta\hat{\epsilon}_{\ell}^{n}-\SH\alpha\,\frac{\partial J}{\partial\delta\hat{\epsilon}_{\ell}}=\delta\hat{\epsilon}_{\ell}^{n}+\alpha\, V\, S_{ij\ell}\,(m_{ij}-V\, S_{ijk}\delta\hat{\epsilon}_{k}^{n}).\label{eq:Iteration}
\end{equation}
Coefficient $\alpha$ is used to adjust the speed of convergence.
Examples of image reconstruction with Landweber algorithm can be found
for instance in \cite{YangMST1999}.

\section{Spatial resolution}

\subsection{Definition}

The spatial resolution can be defined in position as 

\noindent \begin{center}
\emph{the smallest object displacement $\delta\vec{x}$ that produces
a measurable signal variation.}
\par\end{center}

The spatial resolution can also be defined in discrimination as 

\noindent \begin{center}
\emph{the smallest distance $\delta\vec{x}$ between two identical
objects that produce a signal with a measurable difference compared
to the signal they would produce if they were superimposed.}
\par\end{center}

In the first case, as illustrated in Figures~\ref{fig:SpatialResolution}a
and \ref{fig:SpatialResolution}b, the position accuracy is determined
and $\delta\vec{x}$ is then a kind of positioning resolution. In
the second case, as illustrated in Figures~\ref{fig:SpatialResolution}c
and \ref{fig:SpatialResolution}d, the discrimination between two
identical objects is determined and $\delta\vec{x}$ is a kind of
discriminating resolution. In both cases however a measurable signal
variation means a signal variation which is over the noise level.

\subsection{Positioning resolution}

In the case of ECT sensors, which produce as many signals as the square
of the electrode number $N$ when the tensor symmetry is not taken
into account, a variation above the noise level should take into account
all possible outputs. For instance the mean power produced by all
signals should be above the noise power NP in the measurement bandwidth.
The noise power NP corresponds then to the noise equivalent power
(NEP) multiplied by the measurement bandwidth. The detection criterion
can then be written as 
\begin{equation}
\frac{1}{N^{2}}\sum_{i}\sum_{j}(\delta m_{ij})^{2}\geq\mbox{NP}.\label{eq:Detectivity}
\end{equation}

We choose a cylinder of radius $r$ and of length $\ell$ as the object
which displacement by the distance $\delta\vec{x}$ is to be detected
in the sensor sensitive volume. After (\ref{eq:CapacitiveTensorVariation}),
the signal variation $\Delta Q_{i}$ produced on the electrode $i$
due to the presence of the object is
\begin{equation}
\Delta Q_{i}=V_{j}\,\int\Delta\epsilon(\vec{x})\,\vec{\xi}_{i}\cdot\vec{\xi}_{j}\,\d v=V_{j}\,\Delta\epsilon\,\int_{\mathcal{V}}\left(\vec{\xi}_{i}\cdot\vec{\xi}_{j}\right)\,\d v,\label{eq:SignalPosition1}
\end{equation}
where $\Delta\epsilon(\vec{x})$ corresponds to the variation of permittivity
between the situations with and without the object. It is worth noting
that $\Delta\epsilon(\vec{x})$ is zero if the considered position
$\vec{x}$ is not in the object. Therefore the integral in (\ref{eq:SignalPosition1})
can be restricted to the object volume $\mathcal{V}=\pi r^{2}\ell$
as shown in the right hand side of the expression. When displaced
by a distance $\delta\vec{x}$, the signal variation $\Delta Q_{i}'$
induced by the object becomes
\begin{equation}
\Delta Q_{i}'=V_{j}\,\int\Delta\epsilon(\vec{x}+\delta\vec{x})\,\vec{\xi}_{i}\cdot\vec{\xi}_{j}\,\d v.\label{eq:SignalPosition2a}
\end{equation}
Here the volume over which the integral is calculated is slightly
shifted, but at first order, which is correct providing a small displacement
$\delta\vec{x}$, it is possible to rewrite (\ref{eq:SignalPosition2a})
as
\begin{equation}
\Delta Q_{i}'=V_{j}\,\Delta\epsilon\,\int_{\mathcal{V}}\left(\vec{\xi}_{i}\cdot\vec{\xi}_{j}\right)\,\d v+V_{j}\,\Delta\epsilon\,\int_{\mathcal{V}}\delta\vec{x}\cdot\Grad\left(\vec{\xi}_{i}\cdot\vec{\xi}_{j}\right)\,\d v.\label{eq:SignalPosition2b}
\end{equation}
As a consequence the signal variation $\delta Q_{i}=\Delta Q_{i}'-\Delta Q_{i}$
due to the object displacement by $\delta\vec{x}$ is 
\begin{equation}
\delta Q_{i}=V_{j}\,\Delta\epsilon\,\int_{\mathcal{V}}\delta\vec{x}\cdot\Grad\left(\vec{\xi}_{i}\cdot\vec{\xi}_{j}\right)\,\d v.\label{eq:PositioningResolution}
\end{equation}

In (\ref{eq:Detectivity}), all measurement phases are concerned.
For the sake of simplicity, we assume the same applied voltage $V$
for all these measurement phases, hence
\begin{equation}
\delta m_{ij}=V\,\Delta\epsilon\,\int_{\mathcal{V}}\delta\vec{x}\cdot\Grad\left(\vec{\xi}_{i}\cdot\vec{\xi}_{j}\right)\,\d v\label{eq:PositioningMeasurement}
\end{equation}
and, considering (\ref{eq:Detectivity}), the smallest detectable
displacement $\delta\vec{x}$ verifies
\begin{equation}
\frac{1}{N^{2}}\sum_{i}\sum_{j}\left\{ V\,\Delta\epsilon\,\int_{\mathcal{V}}\delta\vec{x}\cdot\Grad\left(\vec{\xi}_{i}\cdot\vec{\xi}_{j}\right)\,\d v\right\} ^{2}\geq\mbox{NP}.\label{eq:PositioningNoise1}
\end{equation}

For a small volume $\mathcal{V}$, the integrand in (\ref{eq:PositioningNoise1})
can be considered as constant over the volume $\mathcal{V}$ leading
to the approximation 
\begin{equation}
\frac{V^{2}\,\Delta\epsilon^{2}\,\mathcal{V}^{2}}{N^{2}}\sum_{i}\sum_{j}\left\{ \delta\vec{x}\cdot\Grad\left(\vec{\xi}_{i}\cdot\vec{\xi}_{j}\right)\right\} ^{2}\geq\mbox{NP}.\label{eq:PositioningNoise2}
\end{equation}

As expected $\delta\vec{x}$ depends on the permittivity variation
$\Delta\epsilon$, on the applied voltage $V$ and on the object volume
$\mathcal{V}$. Thus the larger these parameters, the larger is the
signal and in turn the better is the positioning resolution for a
given noise power. As a consequence an intrinsic positioning resolution
can be defined by normalizing $\delta\vec{x}$ by $\Delta\epsilon$,
$V$ and $\mathcal{V}$. This intrinsic positioning resolution depends
mainly on the sensitivity density $\vec{\xi}_{i}\cdot\vec{\xi}_{j}$
and, to a lesser extent, on the direction which $\delta\vec{x}$ points
to. The positioning resolution is indeed slightly different if $\delta\vec{x}$
is along the $x$-axis or $y$-axis. Figures~\ref{fig:PositioningResolutionDependance}a
and \ref{fig:PositioningResolutionDependance}b shows that difference
for a joined-electrode sensor and a separated-electrode sensor respectively.
In this figure, the amplitude gives the intrinsic positioning resolution
in cubic meter for a 8-electrode 10-cm-radius sensor. The effective
spatial resolution can then be calculated by multiplying the intrinsic
positioning resolution with $\sqrt{\mbox{NP}}/V\,\Delta\epsilon\,\mathcal{V}$.
As an example, if $\sqrt{\mbox{NP}}=1$~pC, $V=100$~V, $\Delta\epsilon=2\epsilon_{0}$
and $r=1$~cm, the positioning resolution is 1.6~cm in the center
of the sensor and can be less than 1~mm at 8.5~cm from the center
of the sensor. Putting aside the slight difference in $\delta x$
direction, the best intrinsic positioning resolution is finally proportional
to
\begin{equation}
\frac{N}{{\displaystyle \sqrt{\sum_{i}\sum_{j}\left|\Grad\left(\vec{\xi}_{i}\cdot\vec{\xi}_{j}\right)\right|^{2}}}}.\label{eq:FinalPositioningResolution}
\end{equation}

Figure~\ref{fig:IntrinsicPositioningResolution} shows the intrinsic
positioning resolution of the ECT systems described in Figure~\ref{fig:ECTSensor}
comprising either 8 joined or separated electrodes. At first glance
the positioning resolution is relatively uniform in the center of
the sensor sensitive volume but becomes drastically dependent on position
close to electrodes especially when the gap between electrodes is
very small.

\subsection{Discriminating resolution}

The discriminating resolution can be treated identically to the positioning
resolution, at least concerning noise level and objects to discern.
After (\ref{eq:CapacitiveTensorVariation}), the signal variation
$\Delta Q_{i}$ produced on electrode $i$ due to the presence of
two identical objects separated by a distance $\delta\vec{x}$ is
\begin{equation}
\Delta Q_{i}=V_{j}\,\int\left[\Delta\epsilon(\vec{x}-\SH\delta\vec{x})+\Delta\epsilon(\vec{x}+\SH\delta\vec{x})\right]\,\vec{\xi}_{i}\cdot\vec{\xi}_{j}\,\d v.\label{eq:SignalDiscriminating1}
\end{equation}
If objects were superimposed, that is to say for a single object producing
twice as much permittivity variation, the signal variation $\Delta Q_{i}'$
would be 
\begin{equation}
\Delta Q_{i}'=V_{j}\,\int2\Delta\epsilon(\vec{x})\,\vec{\xi}_{i}\cdot\vec{\xi}_{j}\,\d v.\label{eq:SignalDiscriminating2}
\end{equation}
As a consequence the signal variation $\delta Q_{i}=\Delta Q_{i}'-\Delta Q_{i}$
between the two situations (\ref{eq:SignalDiscriminating1}) and (\ref{eq:SignalDiscriminating2})
is 
\begin{equation}
\delta Q_{i}=V_{j}\,\int\left[\Delta\epsilon(\vec{x}-\SH\delta\vec{x})-2\Delta\epsilon(\vec{x})+\Delta\epsilon(\vec{x}+\SH\delta\vec{x})\right]\,\vec{\xi}_{i}\cdot\vec{\xi}_{j}\,\d v.\label{eq:DiscriminatingResolution1}
\end{equation}
Providing a sufficiently small displacement $\delta\vec{x}$, integral
in (\ref{eq:DiscriminatingResolution1}) can be calculated over the
object volume $\mathcal{V}$ by introducing the second order approximation
of $\vec{\xi}_{i}\cdot\vec{\xi}_{j}$. One obtains 
\begin{equation}
\delta Q_{i}=\frac{1}{4}\, V_{j}\,\Delta\epsilon\,\int_{\mathcal{V}}\delta\vec{x}\cdot\Grad\left(\delta\vec{x}\cdot\Grad(\vec{\xi}_{i}\cdot\vec{\xi}_{j})\right)\,\d v.\label{eq:DiscriminatingResolution2}
\end{equation}

For the sake of simplicity, we still assume the same applied voltage
$V$ for all measurement phases. Therefore one obtains 
\begin{equation}
\delta m_{ij}=\frac{1}{4}\, V\,\Delta\epsilon\,\int_{\mathcal{V}}\delta\vec{x}\cdot\Grad\left(\delta\vec{x}\cdot\Grad(\vec{\xi}_{i}\cdot\vec{\xi}_{j})\right)\,\d v\label{eq:DiscriminatingMeasurement}
\end{equation}
and, considering (\ref{eq:Detectivity}), the smallest detectable
distance $\delta\vec{x}$ between two similar objects verifies
\begin{equation}
\frac{1}{N^{2}}\sum_{i}\sum_{j}\left\{ \frac{1}{4}\, V\,\Delta\epsilon\,\int_{\mathcal{V}}\delta\vec{x}\cdot\Grad\left(\delta\vec{x}\cdot\Grad(\vec{\xi}_{i}\cdot\vec{\xi}_{j})\right)\,\d v\right\} ^{2}\geq\mbox{NP}.\label{eq:DiscriminatingNoise1}
\end{equation}
For a small volume $\mathcal{V}$, it comes
\begin{equation}
\frac{V^{2}\,\Delta\epsilon^{2}\,\mathcal{V}^{2}}{16N^{2}}\sum_{i}\sum_{j}\left\{ \delta\vec{x}\cdot\Grad\left(\delta\vec{x}\cdot\Grad(\vec{\xi}_{j}\cdot\vec{\xi}_{i})\right)\right\} ^{2}\geq\mbox{NP}.\label{eq:DiscrimnatingNoise2}
\end{equation}

As expected $\delta\vec{x}$ depends on the permittivity variation
$\Delta\epsilon$, on the applied voltage $V$ and on the object volume
$\mathcal{V}$. Thus the larger these parameters, the larger is the
signal and in turn the better is the reachable discriminating resolution
for a given signal to noise ratio. As a consequence an intrinsic discriminating
resolution can be defined by normalizing $\delta\vec{x}$ by $\Delta\epsilon$,
$V$ and $\mathcal{V}$. This intrinsic discriminating resolution
depends mainly on the sensitivity density $\vec{\xi}_{i}\cdot\vec{\xi}_{j}$
and, to a lesser extent, on the direction which $\delta\vec{x}$ points
to. The discriminating resolution is also slightly different if $\delta\vec{x}$
is along $x$-axis or $y$-axis as shown in Figure~\ref{fig:DiscriminatingResolutionDependance}
for a joined-electrode sensor and a separated-electrode sensor. In
this figure, the amplitude gives the intrinsic discriminating resolution
in square meter for a 8-electrode 10-cm-radius sensor. The effective
spatial resolution can then be calculated by multiplying the intrinsic
discriminating resolution with $(\sqrt{\mbox{NP}}/V\,\Delta\epsilon\,\mathcal{V})^{0.5}$.
As an example, if $\sqrt{\mbox{NP}}=1$~pC, $V=100$~V, $\Delta\epsilon=2\epsilon_{0}$
and $r=1$~cm, the discriminating resolution is 4.4~cm in the center
of the sensor and 1~cm at 7.5~cm from the center of the sensor.
Putting aside that slight difference, the best discriminating resolution
is finally proportional to
\begin{equation}
\sqrt{\frac{4N}{\sqrt{{\displaystyle \sum_{i}\sum_{j}\left\{ \Div\left(\Grad(\vec{\xi}_{j}\cdot\vec{\xi}_{i})\right)\right\} ^{2}}}}}.\label{eq:FinalDiscriminatingResolution}
\end{equation}

Figure~\ref{fig:IntrinsicDiscriminatingResolution} shows the intrinsic
discriminating resolution of the ECT sensors described in Figure~\ref{fig:ECTSensor}.
At first glance the discriminating resolution is relatively uniform
in the center of the sensor sensitive volume but becomes drastically
dependent on position close to electrodes especially when the gap
between the electrode is small.

\subsection{Resolution versus sensor diameter, electrode number and gap}

The intrinsic positioning and discriminating resolutions depend on
the sensor geometry through the sensitivity density $\vec{\xi}_{i}\cdot\vec{\xi}_{j}$.
It is then of interest to estimate the influence of the sensor diameter,
the number of its electrodes and the gap between the electrodes.

Concerning the sensor diameter, the sensitivity field $\vec{\xi}_{i}$
is exactly inversely proportional to the sensor radius $R$ and depends
on the normalized coordinate $z/R$ whatever the shape of the electrodes.
Therefore the sensitivity density is exactly inversely proportional
to the square of the sensor radius and any spatial derivative introduces
an additional division by $R$. As a consequence, for a given object,
applied voltage and noise, the positioning resolution is proportional
to the cube of the sensor radius whereas the discriminating resolution
is proportional to the square of the sensor radius. Then increasing
the sensor diameter by a factor 2 increases the smallest detectable
displacement $\delta\vec{x}$ of an object by a factor 8 and the smallest
detectable distance $\delta\vec{x}$ between two similar objects by
a factor 4. In order to compensate that loss, it is required to increase
the applied voltage accordingly, that is to say as the cube of the
sensor radius when considering the positioning resolution and as the
square of the sensor radius when considering the discriminating resolution.

Concerning the number of electrodes and the gap between electrodes,
the calculation is less straightforward. Because resolutions have
the same symmetry as the sensor structure, it is interesting to consider
two specific cross sections, the first from the center of the sensor
($z=0$) to the center of an electrode ($z=R$) and the second from
the center of the sensor ($z=0$) to the center of the gap ($z=R\,\exp(\imath\pi/N)$).

Figure~\ref{fig:ElectrodeDependence} shows the dependence of the
intrinsic positioning (\ref{fig:ElectrodeDependence}a and \ref{fig:ElectrodeDependence}b)
and discriminating (\ref{fig:ElectrodeDependence}c and \ref{fig:ElectrodeDependence}d)
resolutions on the number of separated electrodes with $\eta=50$\%
for a 10-cm-radius sensor along a cross section from the sensor center
to the electrode center (\ref{fig:ElectrodeDependence}a and \ref{fig:ElectrodeDependence}c)
and along the sensor center to the gap center (\ref{fig:ElectrodeDependence}b
and \ref{fig:ElectrodeDependence}d). There is little difference between
the two considered cross sections except for small electrode number.
In that latter case, it can be noticed that positioning and discriminating
resolutions are better in the center of the sensor respectively for
3 and 4 electrodes. When the number of electrodes increases, the difference
between the best and the worst spatial resolution increases since
the resolution increases close to the electrodes and decreases in
the center.

Figure~\ref{fig:GapDependence} shows the dependence of the intrinsic
positioning (\ref{fig:GapDependence}a and \ref{fig:GapDependence}b)
and discriminating (\ref{fig:GapDependence}c and \ref{fig:GapDependence}d)
resolutions on the gap between the electrodes for a 10-cm-radius sensor
along a cross section from the sensor center to the electrode center
(\ref{fig:GapDependence}a and \ref{fig:GapDependence}c) and along
the sensor center to the gap center (\ref{fig:GapDependence}b and
\ref{fig:GapDependence}d). In the center of the sensor, the influence
of the gap is small whereas it becomes very important close to the
electrodes. The difference between the best and the worst resolution
is minimized for a small gap when considering the cross section to
the electrode center whereas it is minimized for large gap when considering
the cross section to the gap center. A good compromise is obtained
for a 50\% gap though the ratio between the worst and the best intrinsic
resolutions is still 15 for the positioning resolution and 10 for
the discriminating resolution.

\subsection{Resolution estimation from simulations}

Expression (\ref{eq:AnalyticSensitivityFieldJoined0}) of the electric
field is an exact solution for a joined electrode sensor. However
it is not the case for the separated-electrode sensor. Expression
(\ref{eq:AnalyticSensitivityFieldSeparated0}) is indeed only an approximation
of the electric field bcause the gaps in the grounded electrode have
been neglected. In order to verify the resolution calculations of
the separated-electrode ECT sensor, it is worth to make numerical
simulations. We used gMsh GetDP softwares \cite{GeuzainePAMM2007,GeuzaineIJNME2009}
for simulations and calculations. The results of (\ref{eq:FinalPositioningResolution})
and (\ref{eq:FinalDiscriminatingResolution}) for a 8-separated-electrode
sensor with 50\% gap are respectively shown in Figures~\ref{fig:IntrinsicSimuledResolution}a
and \ref{fig:IntrinsicSimuledResolution}b. Of course, numerical simulations
exhibit more noise in front of analytic calculations but it can be
clearly concluded that approximation (\ref{eq:AnalyticSensitivityFieldSeparated0})
gives results very close to the simulation ones. The shapes are indeed
similar though resolutions are slightly overestimated in the analytic
calculation with approximation (\ref{eq:AnalyticSensitivityFieldSeparated0}).

The numerical simulations can also be used to study other geometries,
for instance a square cross section instead of a circle. A question
arises then:

\noindent \begin{center}
\emph{Is it better or not to put electrodes in the corners ?}
\par\end{center}

Positioning and discriminating resolutions are shown for these two
situations in Figure~\ref{fig:IntrinsicSimuledSquareResolution}
when considering an 8-electrode sensor with 50\% gap between electrodes
and a grounded guard electrode outside the sensor. It can be seen
that a better resolution is obtained if there are electrodes in the
corners of the square. Moreover the resolution contour levels better
follow the sensor geometry with electrode in the corners.

\section{Conclusion}

In this paper we have calculated the spatial resolution in position
and in discrimination in the case of electrical capacitance tomographic
(ECT) systems. For cylindrical ECT systems, it is shown that (i) the
best resolution is obtained for 3 and 4 electrodes in the sensor center,
that (ii) the resolution greatly increases near the electrodes specially
for large electrode number and that (iii) the resolution is better
uniformly distributed when considering 50\% gap between electrodes.
The expression of intrinsic spatial resolution can also be used to
compare different designs and then to choose the most effective one.
For instance we have shown that a 8-electrode square ECT system with
50\% gap presents a better spatial resolution if there are electrodes
in the corners of the square. This opens the way to an objective comparison
between designs for dedicated applications.

\clearpage{}
\bibliographystyle{unsrt}

\clearpage{}

\section*{Figure Captions}

\begin{figure}[H]
\protect\caption{General description of a capacitive sensor. Illustration of the capacitance
between the electrodes. \label{fig:GeneralCapacitiveSensor}}
\end{figure}

\begin{figure}[H]
\protect\caption{Electric field for an ECT sensor with $N=8$ electrodes distributed
around a cylinder of radius $R=0.1$~m when electrode $j=0$ is held
to 1~V (thick black line) while the other electrodes are grounded
(thin black line). (a) Sensitivity field amplitude in m$^{-1}$ for
joined electrodes. (b) Sensitivity field amplitude in m$^{-1}$ for
separated electrodes with $\eta=50$\%, which corresponds to 50\%
gap and 50\% electrode. \label{fig:ECTSensor}}
\end{figure}

\begin{figure}[H]
\protect\caption{Sensitivity map for an 8-electrode ECT sensor with joined electrodes.
Sensitivity maps $S_{01}$, $S_{02}$, $S_{03}$ and $S_{04}$ are
respectively represented in (a), (b) (c) and (d) sub-figures. Any
other sensitivity maps can be obtained by appropriate rotation and
symmetry of these 4 sensitivity maps. The sensor radius is 10~cm
and sensitivities are expressed in m$^{-2}$. Thick and thin black
lines are alternatively used to distinguish the electrodes. One electrode
is held to voltage $V$, another one is connected to a virtual ground
measurement system producing the signal $V_{m}$ and the remaining
electrodes are grounded. \label{fig:ECTSensitivityMapsJoined}}
\end{figure}
\begin{figure}[H]
\protect\caption{Sensitivity map for an 8-electrode ECT sensor with separated electrodes.
Sensitivity maps $S_{01}$, $S_{02}$, $S_{03}$ and $S_{04}$ are
respectively represented in (a), (b) (c) and (d) sub-figures. Any
other sensitivity maps can be obtained by appropriate rotation and
symmetry of these 4 sensitivity maps. The sensor radius is 10~cm
and sensitivities are expressed in m$^{-2}$. Thick and thin black
lines are alternatively used to distinguish the electrodes. One electrode
is held to voltage $V$, another one is connected to a virtual ground
measurement system producing the signal $V_{m}$ and the remaining
electrodes are grounded. \label{fig:ECTSensitivityMapsDisjointed}}
\end{figure}

\begin{figure}[H]
\protect\caption{Illustration of (a,b) the positioning resolution and (c,d) discriminating
resolution. The distance $\delta\vec{x}$ corresponds to the resolution
if signals produced in (a) and (b), for the positioning resolution,
and in (c) and (d), for the discerning resolution, are different by
the noise level. \label{fig:SpatialResolution}}
\end{figure}

\begin{figure}[H]
\protect\caption{Dependence of the intrinsic positioning resolution on the displacement
direction for a 8-electrode 10-cm-radius sensor in the case of (a)
joined electrodes and (b) separated electrodes with $\eta=50$\%.
\label{fig:PositioningResolutionDependance}}
\end{figure}

\begin{figure}[H]
\protect\caption{Intrinsic positioning resolution for (a) a joined-electrode sensor
and (b) a separated-electrode sensor with $\eta=50$\%. Electrodes
are alternatively represented as thin and thick black lines to better
show their shape.  \label{fig:IntrinsicPositioningResolution}}
\end{figure}

\begin{figure}[H]
\protect\caption{Dependence of the intrinsic discriminating resolution on the displacement
direction for a 8-electrode 10-cm-radius sensor in the case of (a)
joined electrodes and (b) separated electrodes with $\eta=50$\%.
\label{fig:DiscriminatingResolutionDependance}}
\end{figure}

\begin{figure}[H]
\protect\caption{Intrinsic discriminating resolution for (a) a joined-electrode sensor
and (b) a separated-electrode sensor with $\eta=50$\%. Electrodes
are alternatively represented as thin and thick black lines to better
show their shape. \label{fig:IntrinsicDiscriminatingResolution}}
\end{figure}

\begin{figure}[H]
\protect\caption{Dependence of the intrinsic resolution on the number of electrodes
for a 10-cm-radius sensor with $\eta=50$\%. Intrinsic positioning
resolution from the sensor center (a) to the electrode center or (b)
to the gap center. Intrinsic discriminating resolution from the sensor
center (c) to the electrode center or (d) to the gap center. \label{fig:ElectrodeDependence}}
\end{figure}

\begin{figure}[H]
\protect\caption{Dependence of the intrinsic resolution on the gap between electrodes
for a 8-electrode 10-cm-radius sensor. Intrinsic positioning resolution
from the sensor center (a) to the electrode center or (b) to the gap
center. Intrinsic discriminating resolution from the sensor center
(c) to the electrode center or (d) to the gap center. \label{fig:GapDependence}}
\end{figure}

\begin{figure}[H]
\protect\caption{Intrinsic (a) positioning and (b) discriminating resolutions for a
10-cm-radius 8-separated-electrode sensor with $\eta=50$\% calculated
from electrostatic simulations with a finite element method. \label{fig:IntrinsicSimuledResolution}}
\end{figure}

\begin{figure}[H]
\protect\caption{Intrinsic (a, c) positioning and (b, d) discriminating resolutions
for a 20-cm-square 8-electrode sensor with 50\% gap and a grounded
guard electrode whether there are (a, b) or not (c, d) measurement
electrodes in the corners of the square. \label{fig:IntrinsicSimuledSquareResolution}}
\end{figure}

\clearpage{}

\Fig{\ref{fig:GeneralCapacitiveSensor}}{\nameref{fig:GeneralCapacitiveSensor}}

\noindent \begin{center}
\figfont
\psfrag{C12}[Br][Br]{$C_{12}$}
\psfrag{C13}[Br][Br]{$C_{13}$}
\psfrag{C14}[Br][Br]{$C_{14}$}
\psfrag{C23}[Br][Br]{$C_{23}$}
\psfrag{C24}[Br][Br]{$C_{24}$}
\psfrag{C34}[Br][Br]{$C_{34}$}
\psfrag{E1}[Bc][Bc]{Electrode 1}
\psfrag{E2}[Bc][Bc]{Electrode 2}
\psfrag{E3}[Bc][Bc]{Electrode 3}
\psfrag{E4}[Bc][Bc]{Electrode at infinity}
\psfrag{Eps1}[Bc][Bc]{Dielectric 1}
\psfrag{Eps2}[Bc][Bc]{Dielectric 2}
\includegraphics{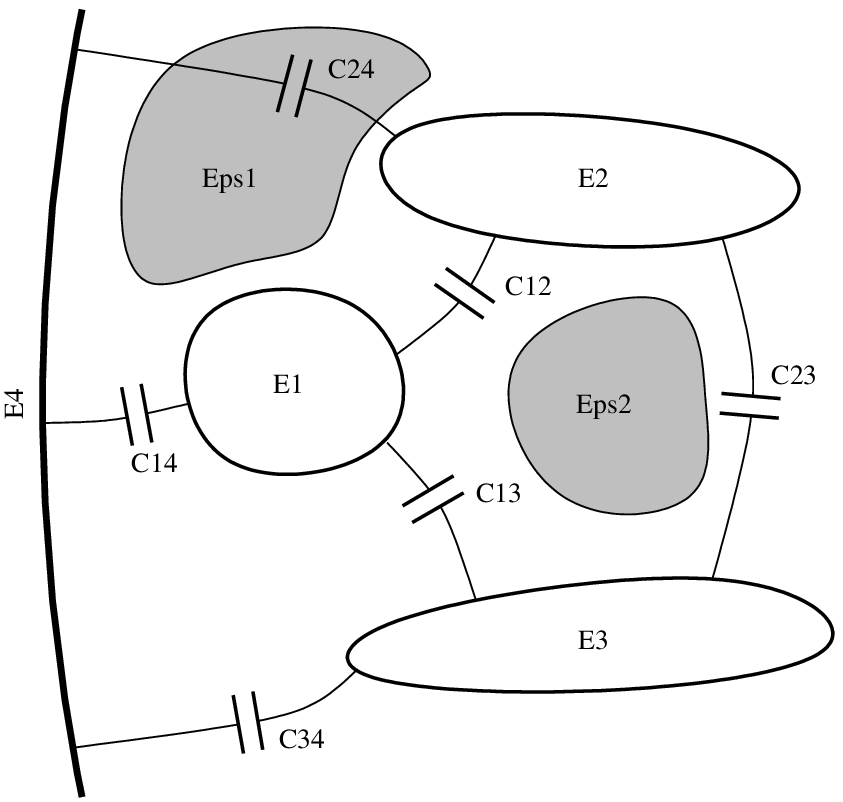}\clearpage{}
\par\end{center}

\Fig{\ref{fig:ECTSensor}}{\nameref{fig:ECTSensor}}

\noindent \begin{center}
\figfont
\psfrag{a}[Br][Br]{(a)}
\psfrag{b}[Br][Br]{(b)}
\psfrag{1V}[Bc][Bc]{1~V}
\includegraphics{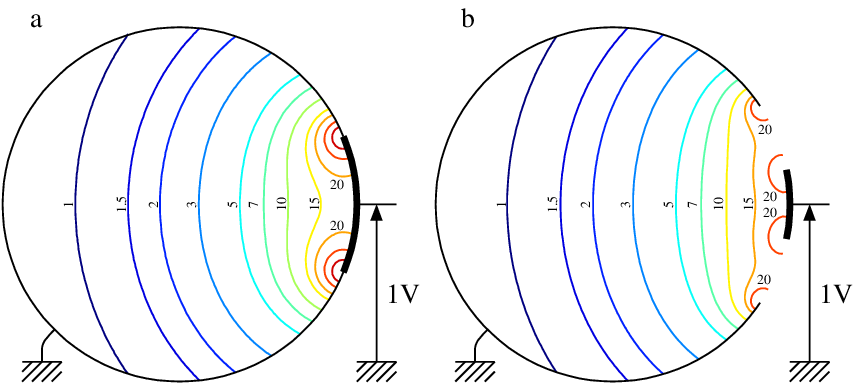}\clearpage{}
\par\end{center}

\Fig{\ref{fig:ECTSensitivityMapsJoined}}{\nameref{fig:ECTSensitivityMapsJoined}}

\noindent \begin{center}
\figfont
\psfrag{a}[Br][Br]{(a)}
\psfrag{S01}[Bc][Bc]{$S_{01}$}
\psfrag{b}[Br][Br]{(b)}
\psfrag{S02}[Bc][Bc]{$S_{02}$}
\psfrag{c}[Br][Br]{(c)}
\psfrag{S03}[Bc][Bc]{$S_{03}$}
\psfrag{d}[Br][Br]{(d)}
\psfrag{S04}[Bc][Bc]{$S_{04}$}
\psfrag{1V}[Bl][Bl]{1~V}
\psfrag{Vm}[Bl][Bl]{$Vm$}
\includegraphics{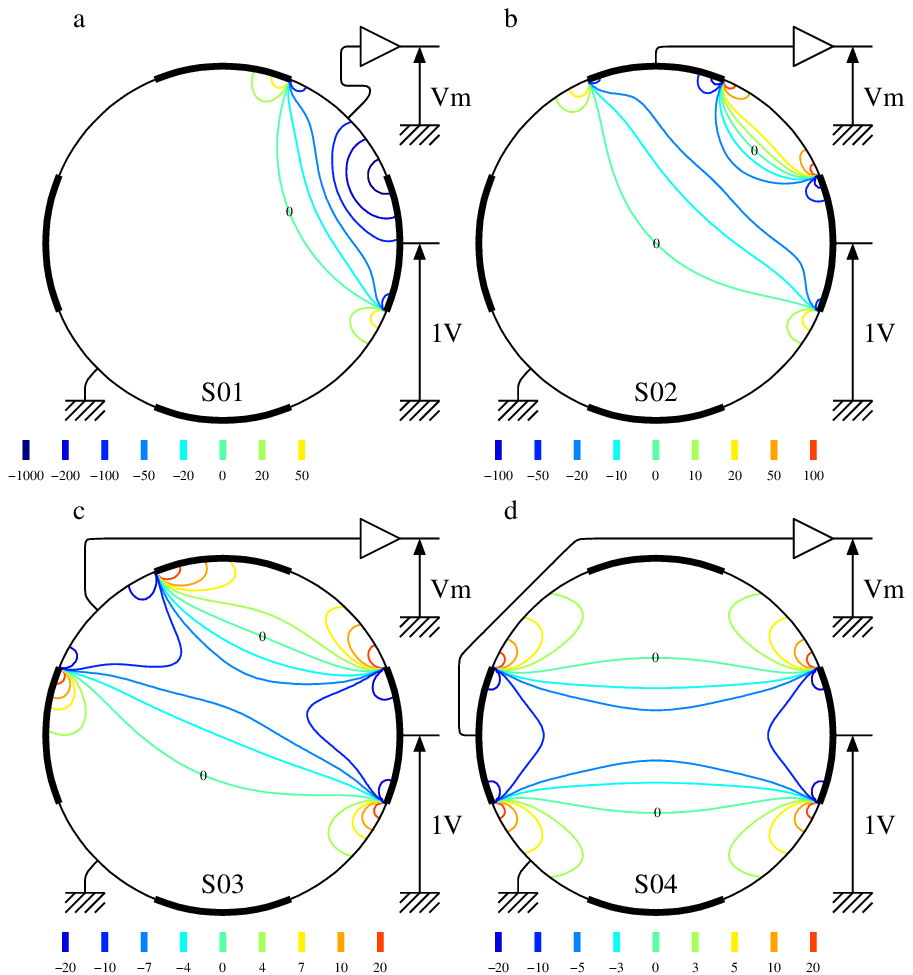}\clearpage{}
\par\end{center}

\Fig{\ref{fig:ECTSensitivityMapsDisjointed}}{\nameref{fig:ECTSensitivityMapsDisjointed}}

\noindent \begin{center}
\figfont
\psfrag{a}[Br][Br]{(a)}
\psfrag{S01}[Bc][Bc]{$S_{01}$}
\psfrag{b}[Br][Br]{(b)}
\psfrag{S02}[Bc][Bc]{$S_{02}$}
\psfrag{c}[Br][Br]{(c)}
\psfrag{S03}[Bc][Bc]{$S_{03}$}
\psfrag{d}[Br][Br]{(d)}
\psfrag{S04}[Bc][Bc]{$S_{04}$}
\psfrag{1V}[Bl][Bl]{1~V}
\psfrag{Vm}[Bl][Bl]{$V_m$}
\includegraphics{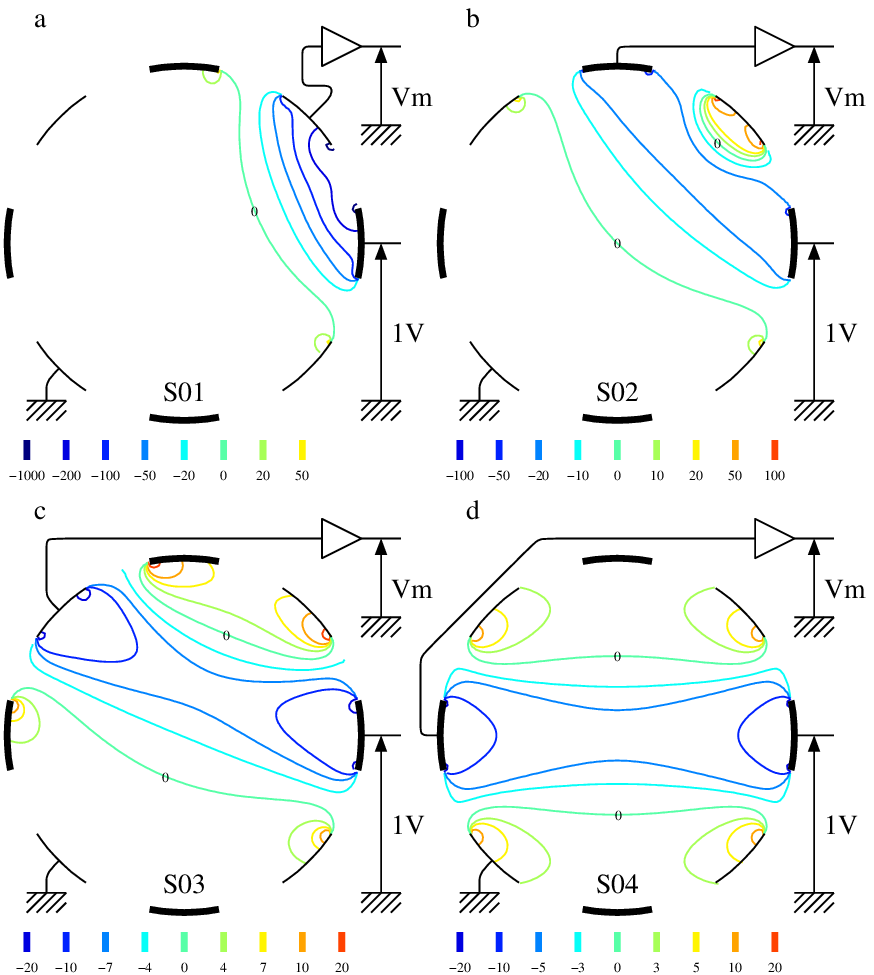}
\par\end{center}

\clearpage{}

\Fig{\ref{fig:SpatialResolution}}{\nameref{fig:SpatialResolution}}

\noindent \begin{center}
\figfont
\psfrag{a}[Br][Br]{(a)}
\psfrag{b}[Br][Br]{(b)}
\psfrag{dx}[Bl][Bl]{$\delta\vec x$}
\psfrag{De}[Bc][Bc]{$\Delta\epsilon$}
\psfrag{c}[Br][Br]{(c)}
\psfrag{d}[Br][Br]{(d)}
\psfrag{2De}[Bc][Bc]{$2\Delta\epsilon$}
\includegraphics{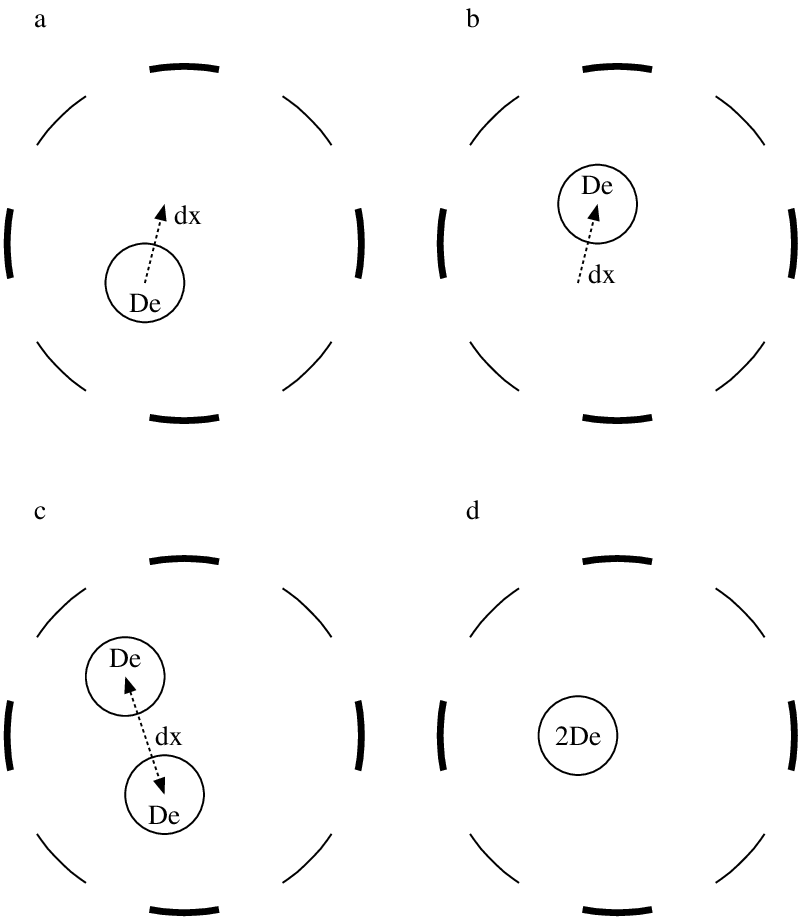}
\par\end{center}

\clearpage{}

\Fig{\ref{fig:PositioningResolutionDependance}}{\nameref{fig:PositioningResolutionDependance}}

\noindent \begin{center}
\figfont
\psfrag{a}[Br][Br]{(a)}
\psfrag{0.0005}[Br][Br]{0.0005}
\psfrag{0.001}[Br][Br]{0.001}
\psfrag{0.002}[Br][Br]{0.002}
\psfrag{0.005}[Br][Br]{0.005}
\psfrag{0.01}[Br][Br]{0.01}
\psfrag{-0.1}[Bc][Bc]{$-10$}
\psfrag{-0.05}[Bc][Bc]{$-5$}
\psfrag{0}[Bc][Bc]{$0$}
\psfrag{+0.05}[Bc][Bc]{$+5$}
\psfrag{+0.1}[Bc][Bc]{$+10$}
\psfrag{b}[Br][Br]{(b)}
\psfrag{Intrinsic dx}[Bc][Bc]{Intrinsic $\delta x$ (m$^{3}$)}
\psfrag{Position}[Bc][Bc]{Position along $x$-axis (cm)}
\psfrag{Along x}[Br][Br]{$\delta\vec{x}$ along $x$-axis}
\psfrag{Along y}[Br][Br]{$\delta\vec{x}$ along $y$-axis}
\includegraphics{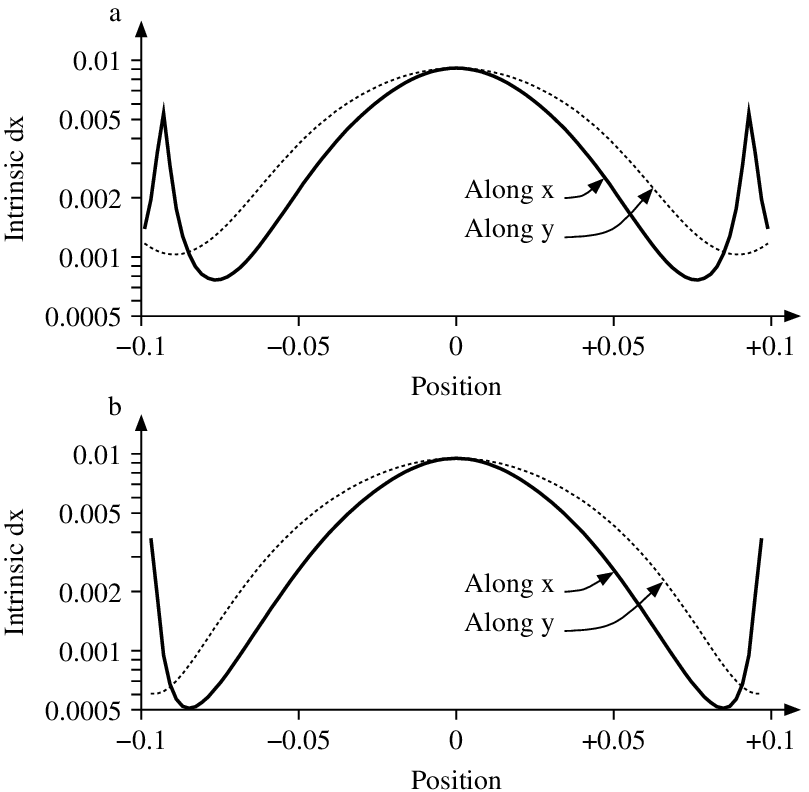}
\par\end{center}

\clearpage{}

\Fig{\ref{fig:IntrinsicPositioningResolution}}{\nameref{fig:IntrinsicPositioningResolution}}

\noindent \begin{center}
\figfont
\psfrag{a}[Br][Br]{(a)}
\psfrag{b}[Br][Br]{(b)}
\psfrag{Intrinsic dx}[Bc][Bc]{$1000\times$intrinsic $\delta x$
(m$^{3}$)}
\includegraphics{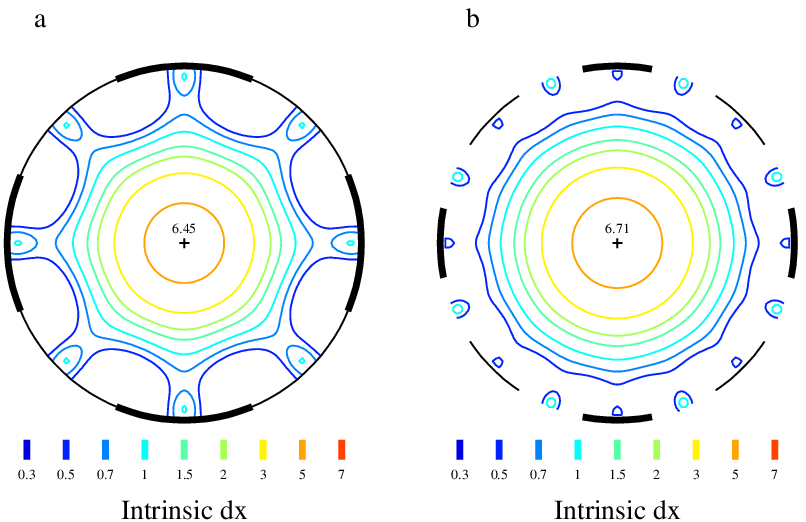}
\par\end{center}

\clearpage{}

\Fig{\ref{fig:DiscriminatingResolutionDependance}}{\nameref{fig:DiscriminatingResolutionDependance}}

\noindent \begin{center}
\figfont
\psfrag{a}[Br][Br]{(a)}
\psfrag{0.002}[Br][Br]{0.002}
\psfrag{0.005}[Br][Br]{0.005}
\psfrag{0.01}[Br][Br]{0.01}
\psfrag{0.02}[Br][Br]{0.02}
\psfrag{0.05}[Br][Br]{0.05}
\psfrag{-0.1}[Bc][Bc]{$-10$}
\psfrag{-0.05}[Bc][Bc]{$-5$}
\psfrag{0}[Bc][Bc]{$0$}
\psfrag{+0.05}[Bc][Bc]{$+5$}
\psfrag{+0.1}[Bc][Bc]{$+10$}
\psfrag{b}[Br][Br]{(b)}
\psfrag{Intrinsic dx}[Bc][Bc]{Intrinsic $\delta x$ (m$^{2}$)}
\psfrag{Position}[Bc][Bc]{Position along $x$-axis (cm)}
\psfrag{Along x}[Br][Br]{$\delta\vec{x}$ along $x$-axis}
\psfrag{Along y}[Br][Br]{$\delta\vec{x}$ along $y$-axis}
\includegraphics{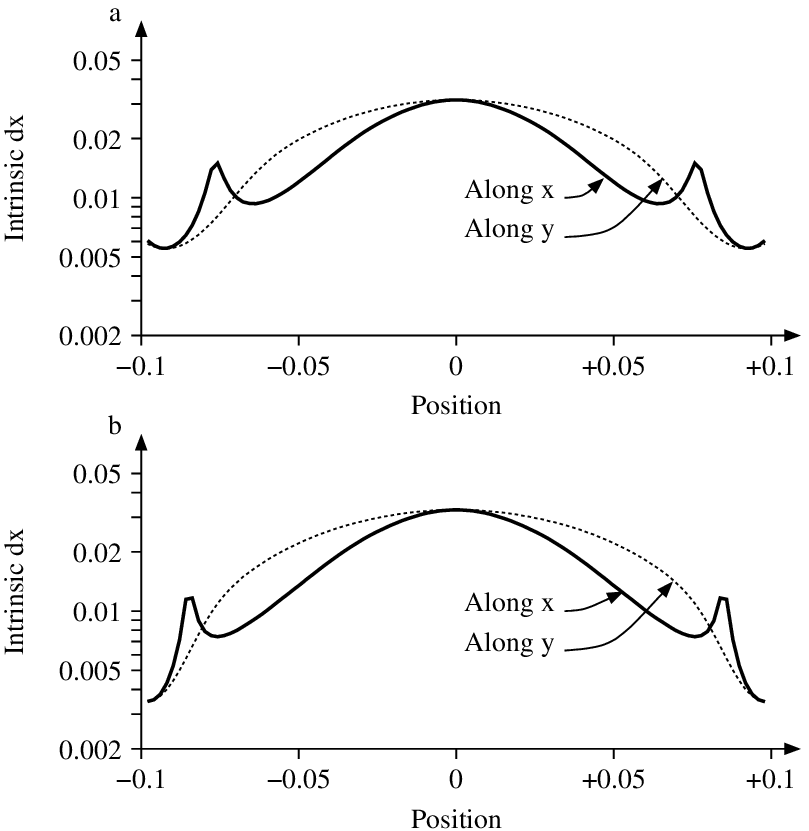}
\par\end{center}

\clearpage{}

\Fig{\ref{fig:IntrinsicDiscriminatingResolution}}{\nameref{fig:IntrinsicDiscriminatingResolution}}

\noindent \begin{center}
\figfont
\psfrag{a}[Br][Br]{(a)}
\psfrag{b}[Br][Br]{(b)}
\psfrag{Intrinsic dx}[Bc][Bc]{$1000\times$intrinsic $\delta x$
(m$^{2}$)}
\includegraphics{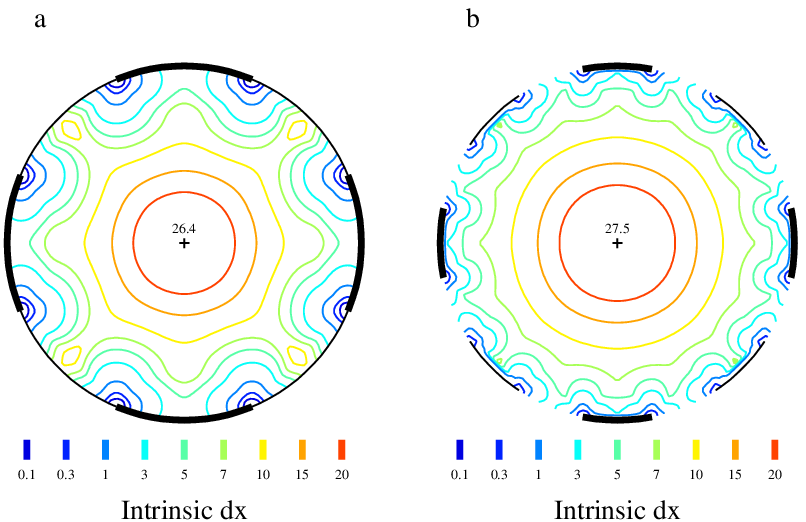}\clearpage{}
\par\end{center}

\Fig{\ref{fig:ElectrodeDependence}}{\nameref{fig:ElectrodeDependence}}

\noindent \begin{center}
\figfont
\psfrag{a}[Br][Br]{(a)}
\psfrag{0.0001}[Br][Br]{$10^{-4}$}
\psfrag{0.001}[Br][Br]{$10^{-3}$}
\psfrag{0.01}[Br][Br]{$10^{-2}$}
\psfrag{0}[Bc][Bc]{0}
\psfrag{2cm}[Bc][Bc]{2}
\psfrag{4cm}[Bc][Bc]{4}
\psfrag{6cm}[Bc][Bc]{6}
\psfrag{8cm}[Bc][Bc]{8}
\psfrag{10cm}[Bc][Bc]{10}
\psfrag{b}[Br][Br]{(b)}
\psfrag{c}[Br][Br]{(c)}
\psfrag{d}[Br][Br]{(d)}
\psfrag{Positioning dx}[Bc][Bc]{Intrinsic $\delta x$ (m$^{3}$)}
\psfrag{Discriminating dx}[Bc][Bc]{Intrinsic $\delta x$ (m$^{2}$)}
\psfrag{Position}[Bc][Bc]{Radial position (cm)}
\includegraphics{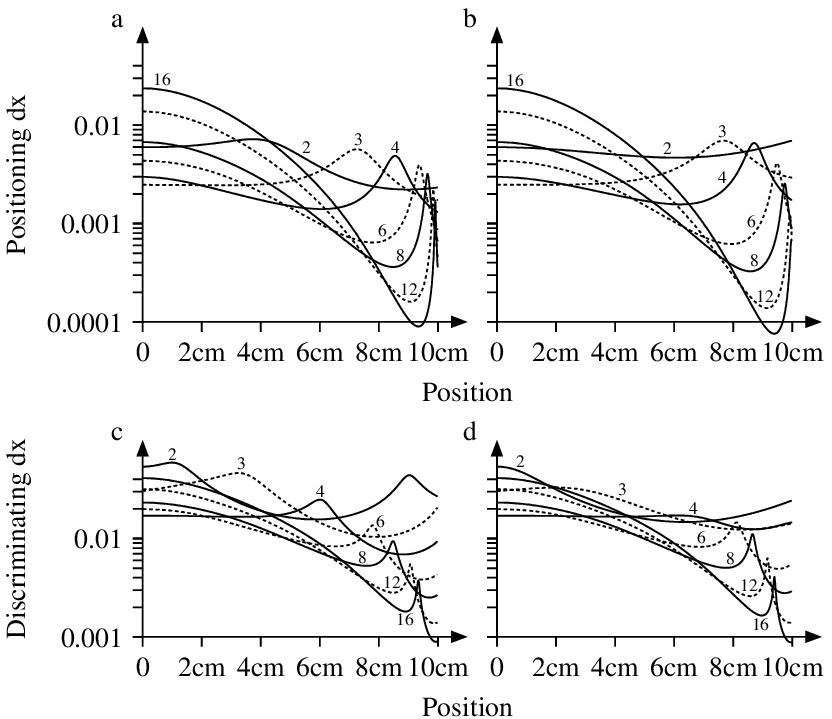}\clearpage{}
\par\end{center}

\Fig{\ref{fig:GapDependence}}{\nameref{fig:GapDependence}}

\noindent \begin{center}
\figfont
\psfrag{a}[Br][Br]{(a)}
\psfrag{-1}[Br][Br]{$10^{-1}$}
\psfrag{-2}[Br][Br]{$10^{-2}$}
\psfrag{-3}[Br][Br]{$10^{-3}$}
\psfrag{-4}[Br][Br]{$10^{-4}$}
\psfrag{-5}[Br][Br]{$10^{-5}$}
\psfrag{-6}[Br][Br]{$10^{-6}$}
\psfrag{0}[Bc][Bc]{0}
\psfrag{2cm}[Bc][Bc]{2}
\psfrag{4cm}[Bc][Bc]{4}
\psfrag{6cm}[Bc][Bc]{6}
\psfrag{8cm}[Bc][Bc]{8}
\psfrag{10cm}[Bc][Bc]{10}
\psfrag{b}[Br][Br]{(b)}
\psfrag{c}[Br][Br]{(c)}
\psfrag{d}[Br][Br]{(d)}
\psfrag{Positioning dx}[Bc][Bc]{Intrinsic $\delta x$ (m$^{3}$)}
\psfrag{Discriminating dx}[Bc][Bc]{Intrinsic $\delta x$ (m$^{2}$)}
\psfrag{Position}[Bc][Bc]{Radial position (cm)}
\includegraphics{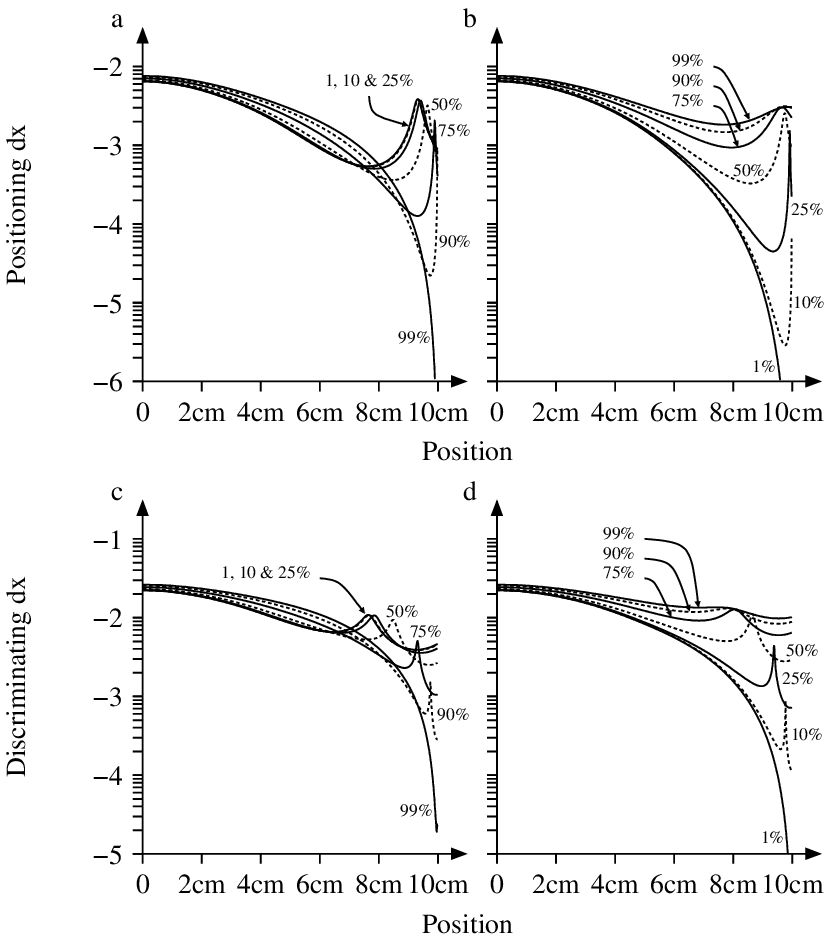}
\par\end{center}

\clearpage{}

\Fig{\ref{fig:IntrinsicSimuledResolution}}{\nameref{fig:IntrinsicSimuledResolution}}

\noindent \begin{center}
\figfont
\psfrag{a}[Br][Br]{(a)}
\psfrag{b}[Br][Br]{(b)}
\psfrag{Intrinsic pos}[Bc][Bc]{$1000\times$intrinsic $\delta x$
(m$^{3}$)}
\psfrag{Intrinsic dis}[Bc][Bc]{$1000\times$intrinsic $\delta x$
(m$^{2}$)}
\includegraphics{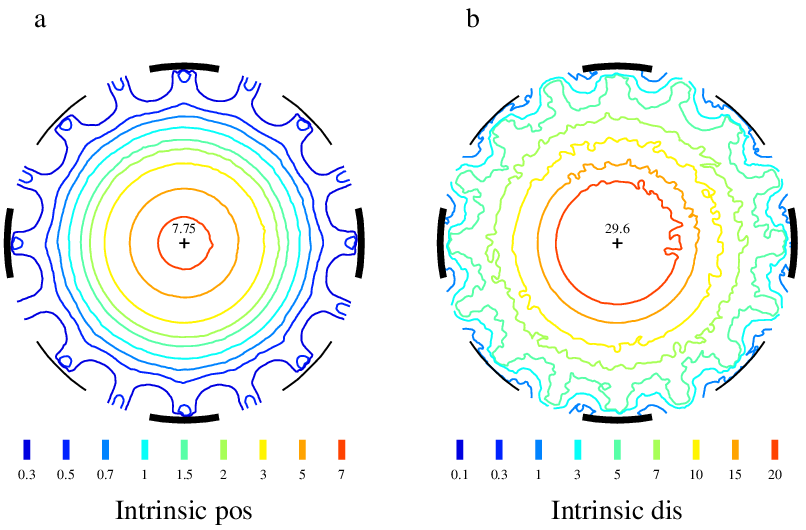}
\par\end{center}

\clearpage{}

\Fig{\ref{fig:IntrinsicSimuledSquareResolution}}{\nameref{fig:IntrinsicSimuledSquareResolution}}

\noindent \begin{center}
\figfont
\psfrag{a}[Br][Br]{(a)}
\psfrag{b}[Br][Br]{(b)}
\psfrag{c}[Br][Br]{(c)}
\psfrag{d}[Br][Br]{(d)}
\psfrag{Intrinsic pos}[Bc][Bc]{$1000\times$intrinsic $\delta x$
(m$^{3}$)}
\psfrag{Intrinsic dis}[Bc][Bc]{$1000\times$intrinsic $\delta x$
(m$^{2}$)}
\includegraphics{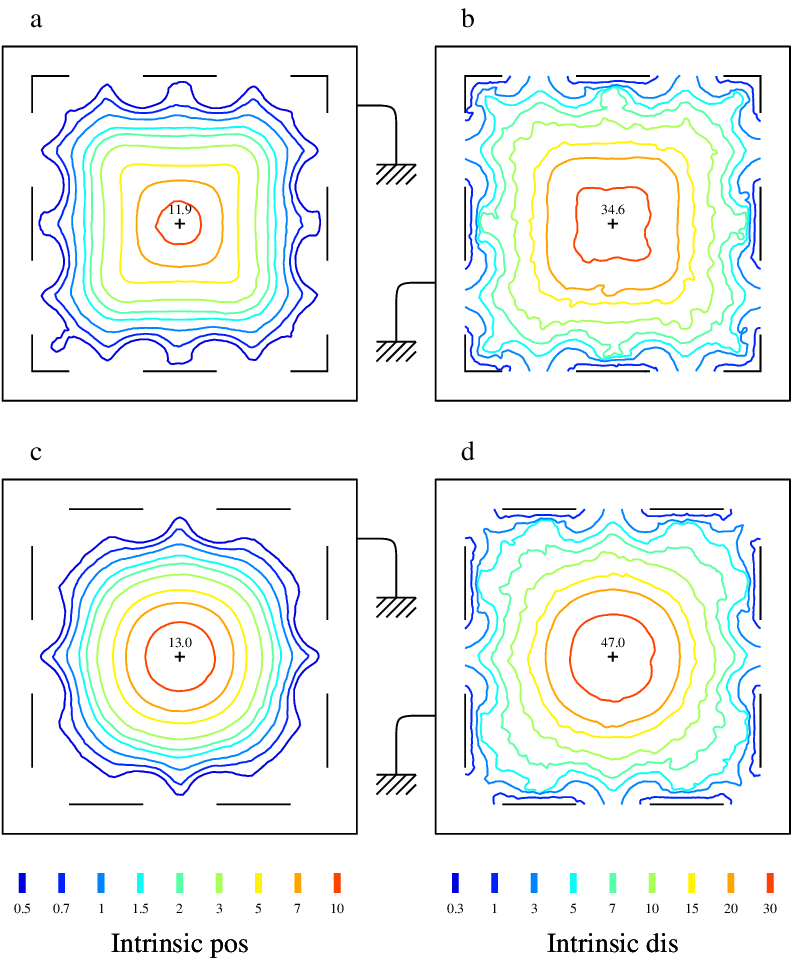}
\par\end{center}
\end{document}